\documentclass[aps,superscriptaddress]{revtex4}
\usepackage{amssymb,amsmath,epsfig}
\usepackage{subfigure}
\usepackage[colorlinks=true, pdfstartview=FitV, linkcolor=blue, citecolor=red, urlcolor=magenta, breaklinks=true]{hyperref}
\begin{document}
\title{Quantum-corrected scattering and absorption of a Schwarzschild black hole with GUP}

\author{M. A. Anacleto}\email{anacleto@df.ufcg.edu.br}
\affiliation{Departamento de F\'{\i}sica, Universidade Federal de Campina Grande
Caixa Postal 10071, 58429-900 Campina Grande, Para\'{\i}ba, Brazil}

\author{F. A. Brito}\email{fabrito@df.ufcg.edu.br}
\affiliation{Departamento de F\'{\i}sica, Universidade Federal de Campina Grande
Caixa Postal 10071, 58429-900 Campina Grande, Para\'{\i}ba, Brazil}
\affiliation{Departamento de F\'isica, Universidade Federal da Para\'iba, 
Caixa Postal 5008, 58051-970 Jo\~ao Pessoa, Para\'iba, Brazil}
 
\author{J. A. V. Campos}\email{joseandrecampos@gmail.com}
\affiliation{Departamento de F\'isica, Universidade Federal da Para\'iba, 
Caixa Postal 5008, 58051-970 Jo\~ao Pessoa, Para\'iba, Brazil}

\author{E. Passos}\email{passos@df.ufcg.edu.br}
\affiliation{Departamento de F\'{\i}sica, Universidade Federal de Campina Grande
Caixa Postal 10071, 58429-900 Campina Grande, Para\'{\i}ba, Brazil}

\begin{abstract}  
In this paper we have implemented quantum corrections for the Schwarzschild black hole metric using the generalized uncertainty principle (GUP) in order to investigate the scattering process.
We mainly compute, at the low energy limit, the differential scattering and absorption cross section by using the partial wave method.
We determine the phase shift analytically and verify that these quantities are modified by the GUP.
We found that due to the quantum corrections from the GUP the absorption is not zero as the mass parameter goes to zero. 
A numerical analysis has also been performed for arbitrary frequencies.

\end{abstract}

\maketitle
\pretolerance10000

\section{Introduction}
The study related to the process of scattering particles through various types of black holes has been a very active field of investigation for strong gravitational fields.
Thus, in black hole physics, its properties are analyzed more efficiently, by investigating the scattering of waves of matter over black holes.
Several studies have been proposed in order to investigate many aspects of the scattering process of scalar waves by black holes in the low energy limit~\cite{Futterman1988,Matzner1977,Westervelt1971,Peters1976,Sanchez1976,Logi1977,
Doram2002,Dolan:2007ut,Crispino:2009ki,Churilov1974,Gibbons1975,Page1976,Churilov1973,Moura:2011rr,
Jung2004,Doran2005,Dolanprd2006,Castineiras2007,Benone:2014qaa,Marinho:2016ixt,Das:1996we,deOliveira:2018kcq,Hai:2013ara}. 
The first works carrying out a numerical investigation of the scattering of planar waves by black holes were examined by Sanchez in the 1970s applied to the Schwarzschild black hole~\cite{Sanchez1978, NSanchez1978}.
General relativity predicts the existence of singularities.  In particular, black hole solutions are often plagued by singularities. 
However, these singularities are generally expected to be found in the inner region of a black hole. 
On the other hand, there are black hole solutions in the literature that have no singularities and are called regular black holes.
The first regular black hole solution was obtained by Bardeen in 1968~\cite{Bardeen} (for a review of regular  black holes see~\cite{Ansoldi:2008jw}).
In the literature there are some works that have explored the scattering processes 
of regular black holes~\cite{Huang:2014nka,Macedo:2015qma,Huang:2015oua,Macedo:2016yyo,konoplya}
in attempting to build a consistent theory of gravity free of singularities or black hole information paradox. 
In recent years, it has been investigated that this type of difficulties can be avoided by considering astrophysical objects with effective radius greater than horizon \cite{nozari}. 
%This increase in the radius of the event horizon is related to the appearance of horizonless compact objects.
For these types of objects we can, for example, mention the fuzzball paradigm~\cite{Mathur:2005zp} proposed in string theory, gravastar~\cite{Mazur:2001fv,Beltracchi:2018ait} and the firewall~\cite{Braunstein:2009my,Almheiri:2012rt}.
By considering quantum corrections of black hole event horizons arising from the generalized uncertainty principle (GUP) \cite{Buoninfante:2020cqz,vag1,vag2,vag3} implies the formation of new types of horizonless objects, that is, the effective radius is greater than the radius of the Schwarzschild horizon. These new objects have been called {\it GUP stars}.
Another example that aims to reconcile general relativity and quantum mechanics is the quantum geometric theory called loop quantum gravity~\cite{Ashtekar:2004eh,Han:2005km,Thiemann:1992jj,Ashtekar:2008zu,Bojowald:2003uh}.
Thus, by considering a semi-classical analysis of loop quantum gravity, a metric with quantum corrections for a self-dual black hole has been obtained in the literature~\cite{Modesto:2008im,Modesto:2009ve}.
In~\cite{Silva:2012mt} the thermodynamics of the self-dual black hole was investigated through tunneling formalism and in~\cite{Anacleto:2015mma} this analyzes was extended by considering the GUP. We have very recently explored the scattering of scalar waves through a self-dual black hole and found that due to the contribution of the minimum area the absorption is different from zero when the mass parameter tends to zero~\cite{Anacleto:2020zhp}. 
In this work we will verify whether this effect remains or not when we consider the effect of quantum corrections due to GUP on the scattering process for a Schwarzschild black hole.

Recently, interest in the study of scattering in gravitational theories has increased, mainly due to the observation of gravitational waves detected by the LIGO/Virgo collaboration~\cite{Abbott:2016blz,TheLIGOScientific:2017qsa}.
The purpose of this work is to explore the effect of quantum gravity corrections that contribute 
to the process of scattering and absorption of particles by a Schwarzschild black hole with GUP.
In order to compute the absorption and the differential cross section we will apply the partial wave method and use the technique introduced in~\cite{Anacleto:2017kmg}.
Thus, we will determine analytically the phase shift at the low energy limit. 
We have shown that by increasing the values assigned to the GUP parameter the absorption/differential cross section is decreased when a {\it linear and quadratic GUP} is considered.
On the other hand, assuming only the {\it quadratic GUP} the absorption/differential cross section is increased.
In addition, we have obtained that in the limit as the mass parameter goes to zero the absorption does not vanish due to the contribution of the GUP. 
Moreover, the analysis for the high frequency regime has been investigated by numerically solving the radial equation for arbitrary frequencies.

The paper is organized as follows. In Sec.~\ref{qcmet} we derive the phase shift and calculate the differential scattering/absorption cross section for a Schwarzschild black hole with GUP by considering analytical and numerical analysis. In Sec.~\ref{conc} we make our final considerations. 

\section{Quantum-corrections to the metric}
\label{qcmet}
Here in order to implement quantum corrections for the Schwarzschild black hole metric and investigating the scattering process, we begin our analysis by considering the generalized uncertainty principle (GUP)~\cite{ADV, Tawfik:2014zca, KMM, Tawfik:2015kga, Gangopadhyay:2015zma}, 
which has been defined as
\begin{eqnarray}
\label{gup}
\Delta x\Delta p\geq \frac{\hbar}{2}\left( 1-\frac{\alpha l_p}{\hbar} \Delta p +\frac{\alpha^2 l^2_p}{\hbar^2} (\Delta p)^2 \right),
\end{eqnarray}
where $\alpha$ is a dimensionless positive parameter and $ l_p $ is the Planck length.

Now the equation (\ref{gup}) can be recast in the form
\begin{eqnarray}
\Delta p\geq \frac{\hbar(2\Delta x +\alpha l_p)}{2\alpha^2 l_p^2}
\left(1- \sqrt{1-\frac{4\alpha^2 l_p^2}{(2\Delta x +\alpha l_p)^2}}\right).
\end{eqnarray}
%Taking into account that $ l_p\ll 1$ and so without loss of generality 
In the following steps and without loss of generality, we shall adopt the units $ G=c=k_B=\hbar=l_p=1 $. So, performing a power series in $\alpha$ we obtain
\begin{eqnarray}
\label{p}
\Delta p\geq \frac{1}{2\Delta x}\left[1-\frac{\alpha}{2\Delta x}+ \frac{\alpha^2}{2(\Delta x)^2}+\cdots    \right].
\end{eqnarray}
For the case $ \alpha=0 $ we recover the Heisenberg uncertainty principle
\begin{eqnarray}
\label{rhup}
\Delta x\Delta p\geq \frac{1}{2}.
\end{eqnarray}
Therefore, from (\ref{rhup}) we can obtain a bound for massless particles that is given by
\begin{eqnarray}
E\Delta x\geq \frac{1}{2}.
\end{eqnarray}
In this case, the equation (\ref{p}) can be written as follows
\begin{eqnarray}
\label{rdgup}
{\cal E}\geq E\left[1-\frac{\alpha}{2(\Delta x)}+ \frac{\alpha^2}{2(\Delta x)^2}+\cdots    \right].
\end{eqnarray} 
The relation (\ref{rdgup}) can still be written in terms of the mass assuming that  $\Delta p\sim E\sim M$ and $ \Delta x\sim r_h=2M $ 
and thus we obtain the following dispersion relation
\begin{eqnarray}
{\cal M}=M_{gup}\geq M\left( 1- \frac{\alpha}{4M}+\frac{\alpha^2}{8M^2} \right).
\end{eqnarray}
From the expression above we can obtain a relation to the event horizon
\begin{eqnarray}
{r}_{hgup} = {2M_{gup}}\geq r_h\left( 1- \frac{\alpha}{2r_h}+\frac{\alpha^2}{2r^2_h} \right).
\end{eqnarray}
When considering only the quadratic GUP, the above relationship becomes
\begin{eqnarray}
{r}_{hgup} = {2M_{gup}}\geq r_h\left(1 +\frac{\alpha^2}{2r^2_h} \right).
\end{eqnarray}
Note that in this case the event horizon due to the GUP is now greater than the event horizon of the Schwarzschild black hole.
Thus the effect of the quadratic GUP naturally implies obtaining quantum astrophysical objects that are called {\it horizonless objects}.

%\subsection{Schwarzschild Black Hole with GUP}
\label{sc2} 
The metric for the Schwarzschild black hole with quantum corrections introduced by the GUP is obtained by replacing the mass $ M $ with the mass GUP $ M_{gup} $ (or replacing the Schwarzschild horizon radius, $ r_h $, with $ r_{hgup} $)
and is described by the following line element
\begin{eqnarray}
\label{metrsd}
ds^2=\left(1-\frac{2M_{gup}}{r}\right) dt^2-\left(1-\frac{2M_{gup}}{r}\right)^{-1}dr^2-r^2\left(d\theta^2 + \sin^2\theta d\phi^2 \right).
\end{eqnarray}
Here the event horizon, $ {r}_{hgup} $, is given by 
\begin{eqnarray}
{r}_{hgup} = {2M_{gup}}=r_h\left( 1- \frac{\alpha}{4M}+\frac{\alpha^2}{8M^2} \right),
\end{eqnarray}
where $ r_h=2M $ is the event horizon of the Schwarzschild black hole.

Note that, taking the limit of $ M\rightarrow 0 $ we find that the radius of the horizon, $ r_ {hgup} $, is not zero, that is
\begin{eqnarray}
{r}_{hgup}\approx\frac{\alpha^2}{4M},
\end{eqnarray}
and in terms of the mass we have
\begin{eqnarray}
{M}_{gup}\approx\frac{\alpha^2}{8M}.
\end{eqnarray}

\subsection{Differential Scattering Cross Section}
In this section we introduce the Schwarzschild black hole with quantum corrections implemented by the GUP for the purpose of determining the differential scattering cross section for this model. 
Thus, for this purpose we adopt the partial wave method to calculate the phase shift at the low energy limit.
We now consider the case of the massless scalar field equation to describe the scattered wave in the background (\ref{metrsd}),  given by 
\begin{eqnarray}
\dfrac{1}{\sqrt{-g}}\partial_{\mu}\Big(\sqrt{-g}g^{\mu\nu}\partial_{\nu}\Psi\Big)=0 .
\end{eqnarray}
By applying a separation of variables into the equation above
\begin{eqnarray}
\Psi_{\omega l m}({\bf r},t)=\frac{{\cal R}_{\omega l}(r)}{r}Y_{lm}(\theta,\phi)e^{-i\omega t},
\end{eqnarray}
where $Y_{lm}(\theta,\phi)  $ are the spherical harmonics and $ \omega $ is the frequency, the radial equation for $ {\cal R}_{\omega l}(r) $ is now written in the following form  
\begin{eqnarray}
\label{eqrad}
\lambda(r)\dfrac{d}{dr}\left(\lambda(r)\dfrac{d{\cal R}_{\omega l}(r)}{dr} \right) +\left[ \omega^2 -V_{eff} \right]{\cal R}_{\omega l}(r)=0,
\end{eqnarray}
where 
\begin{eqnarray}
\lambda(r)=1-\frac{2M_{gup}}{r}=1-\frac{2M}{r}\left( 1- \frac{\alpha}{4M}+\frac{\alpha^2}{8M^2} \right),
\end{eqnarray}
and
\begin{eqnarray}
V_{eff}=\frac{\lambda(r)}{r}\frac{d\lambda(r)}{dr} + \frac{\lambda(r)l(l+1)}{r^2},
\end{eqnarray}
being defined as the effective potential.

Next, in order to write equation (\ref{eqrad}) in the form of a Schroedinger-type equation, we consider a new  change of variables, i.e., the radial function $ \chi(r)=\lambda^{1/2}(r){\cal R}(r) $, such that
\begin{eqnarray}
\label{eqradpsi}
\dfrac{d^2\chi(r)}{dr^2}+V(r) \chi(r) = 0,
\end{eqnarray}
with $ V(r) $ being defined as follows
\begin{eqnarray}
\label{poteff}
V(r)=\dfrac{[\lambda'(r)]^2}{4 \lambda^2(r)} - \dfrac{\lambda''(r)}{2\lambda(r)} + \dfrac{\omega^2}{\lambda^2(r)} - 
\dfrac{V_{eff}}{\lambda^2(r)}.
\end{eqnarray}

At this point we consider equation (\ref{eqradpsi}) and for the potential
$V(r) $ we will perform a power series expansion at $ 1/r $, so that we have
\begin{eqnarray}
\frac{d^2\chi(r)}{dr^2}+\left[\omega^2+ {\cal U}_{eff}(r)\right] \chi(r) = 0,
\end{eqnarray}
where $ U_{eff} $ is a new effective potential given by
\begin{eqnarray}
\label{pot1}
{\cal U}_{eff}(r)= \frac{4 M_{gup}\,\omega^2}{r}+\frac{12\ell^2}{r^2}+\cdots.
\end{eqnarray}
Notice that due to the modification of the term in $1/r^2 $, $ \ell^2 $ has been defined as follows~\cite{Anacleto:2017kmg,Anacleto:2019tdj}: 
\begin{eqnarray}
\label{ell}
\ell^2\equiv-\frac{(l^2+l)}{12}+M^2\omega^2\left( 1-\frac{\alpha}{4M}+\frac{\alpha^2}{8M^2} \right)^2.
\end{eqnarray}
In the limit as $ r\rightarrow 0 $ we have $ {\cal U}_{eff}(r) \rightarrow 0 $ and so the suitable asymptotic behavior is satisfied.

Now we will determine the phase shift analytically, at the low frequency limit, by using the following approximation formula 
\begin{eqnarray}
\label{formapprox}
\delta_l\approx 2(l-\ell)=2\left(l - \sqrt{-\frac{(l^2+l)}{12}
+M^2\omega^2\left( 1-\frac{\alpha}{4M}+\frac{\alpha^2}{8M^2} \right)^2}\right), 
\end{eqnarray}
and since at the low frequency limit the phase shift $ \delta_l $ is obtained by considering $ l\rightarrow 0$,  we have
\begin{eqnarray}
\label{phase2}
\delta_l=-2M\omega\left( 1-\frac{\alpha}{4M}+\frac{\alpha^2}{8M^2} \right)+{\cal O}(l).
\end{eqnarray}
Next, using the expression (\ref{phase2}) for the phase shift, we can obtain the differential scattering cross section by applying the following equation~\cite{Yennie1954,Cotaescu:2014jca}:
\begin{eqnarray}
\label{espalh}
\dfrac{d\sigma}{d\theta}=\big|f(\theta) \big|^2=\Big| \frac{1}{2i{\omega}}\sum_{l=0}^{\infty}(2l+1)\left(e^{2i\delta_l} -1 \right)
\frac{P_{l}\cos\theta}{1-\cos\theta}\Big|^2.
\end{eqnarray}
Now taking the small angle limit the above equation becomes
\begin{eqnarray}
\label{espalh2}
\frac{d\sigma}{d\theta}&=&\frac{4}{\omega^2\theta^4}\Big|\sum_{l=0}^{\infty}(2l+1)\sin(\delta_{l})
{P_{l}\cos\theta}\Big|^2,
\\
&=&\frac{16M^2}{\theta^4}\left( 1-\frac{\alpha}{4M}+\frac{\alpha^2}{8M^2} \right)^2\Big|\sum_{l=0}^{\infty}(2l+1)
{P_{l}\cos\theta}\Big|^2.
\label{espalh3}
\end{eqnarray}
Therefore, at the low frequency limit, that is for $ l = 0 $, we obtain the following result:
\begin{eqnarray}
\label{diffcsect}
\frac{d\sigma}{d\theta}\Big |^{\mathrm{l f}}_{\omega\rightarrow 0}
&=&\frac{16M^2}{\theta^4}\left( 1-\frac{\alpha}{4M}+\frac{\alpha^2}{8M^2} \right)^2+\cdots,
\\
&=&\frac{16}{\theta^4}\left( M^2-\frac{\alpha M}{2}+\frac{5\alpha^2}{16} - \frac{\alpha^3}{16M}+\frac{\alpha^4}{64M^2}\right)+ \cdots.
\end{eqnarray}
Note that for $ \alpha = 0 $ the result for the Schwarzschild black hole is obtained. 
Hence, we verify that the result for the differential scattering cross section of the Schwarzschild black hole with GUP is decreased as we increase the values of the $\alpha$ parameter.
By considering only the {\it quadratic GUP}, the differential scattering cross section becomes
\begin{eqnarray}
\label{diffcsect2}
\frac{d\sigma}{d\theta}\Big |^{\mathrm{l f}}_{\omega\rightarrow 0}
&=&\frac{16M^2}{\theta^4}\left( 1+\frac{\alpha^2}{8M^2} \right)^2+\cdots,
\\
&=&\frac{16}{\theta^4}\left( M^2+\frac{\alpha^2}{4}+\frac{\alpha^4}{64M^2}\right)+\cdots.
\end{eqnarray}
At the limit of $ M\rightarrow 0$ the dominant term of equation (\ref{diffcsect}) becomes nonzero and is given by
\begin{eqnarray}
\label{diffcsect2}
\frac{d\sigma}{d\theta}\Big |^{\mathrm{l f}}_{M\rightarrow 0}
\approx\frac{\alpha^4}{4\theta^4 M^2}.
\end{eqnarray}

\subsection{Absorption Cross Section}

We will now determine the absorption cross section for a Schwarzschild black hole with quantum corrections originating from the GUP at the low frequency  limit.
The total absorption cross section can be determined as following:
\begin{eqnarray}
\label{abscsec}
\sigma_{abs}%=\frac{\pi}{\omega^2}\sum_{l=0}^{\infty}(2l+1)\Big(1-\big|e^{2i\delta_l}\big|^2\Big)
=\frac{\pi}{\omega^2}\sum_{l=0}^{\infty}(2l+1)\Big(\big|1-e^{2i\delta_l}\big|^2\Big)
=\frac{4\pi}{\omega^2}\sum_{l=0}^{\infty}(2l+1)\sin^2(\delta_{l}).
\end{eqnarray}
By considering the low energy limit ($ \omega\rightarrow 0 $) and replacing (\ref{phase2}) in (\ref{abscsec}), the absorption for $ l=0 $ is given by
\begin{eqnarray}
\label{abs1}
\sigma_{abs}^{\mathrm{l f}}
&=& 16\pi{M^2}\left( 1-\frac{\alpha}{4M}+\frac{\alpha^2}{8M^2} \right)^2,
\\
&=&{16\pi}\left( M^2-\frac{\alpha M}{2}+\frac{5\alpha^2}{16} - \frac{\alpha^3}{16M}+\frac{\alpha^4}{64M^2}\right)
=4\pi{r}^2_{hgup}={A}_{schwgup}.
\end{eqnarray}
In the absence of the GUP, when $ \alpha = 0 $,
we recover the result for the absorption of the Schwarzschild black hole.
So we can observe that, by considering the {\it linear and quadratic GUP}, the absorption amplitude has its value reduced when we increase the value of $\alpha$.
However, by assuming only the {\it quadratic GUP}, we find the following result for absorption at the low frequency limit
\begin{eqnarray}
\label{abs2q}
\sigma_{abs}^{\mathrm{l f}}
&=& 16\pi{M^2}\left(1+\frac{\alpha^2}{8M^2} \right)^2,
\\
&=&{16}\left( M^2+\frac{\alpha^2}{4}+\frac{\alpha^4}{64M^2}\right)
\label{abs2qa}.
\end{eqnarray}
In this case the absorption increases when we increase the value of $\alpha$.
From equations (\ref{abs1}) and (\ref{abs2qa}) we can obtain a very interesting result in the limit as the mass goes to zero. In this limit the absorption presents a non-zero value, contrary to the usual case of the Schwarzschild black hole, which is given by
\begin{eqnarray}
\label{abs1a}
\sigma_{abs}^{\mathrm{l f}}
&\approx &\frac{\pi \alpha^4}{4M^2}.
\end{eqnarray}
It is worth noting that, contrarily to the usual case of a Schwarzschild black hole, the 
differential scattering/absorption cross section of a Schwarzschild black hole with quantum corrections implemented by the GUP is different from zero when the mass goes to zero.

\subsection{Numerical analyses}
At this point we show the numerical results that were obtained by numerically solving the radial equation (\ref{eqrad}). For this we follow the same numerical procedure performed in~\cite{Dolan:2012yc}.
The table~\ref{tab1} shows the comparison between the analytical and numerical results of the absorption with {the linear and quadratic GUP} for values of $\alpha$ between 0 and 1, fixing the values of $ M = 1 $ and $ l = 0 $. 
In table~\ref{tab2} we show the comparison between the analytical and numerical results of the absorption with only {\it quadratic GUP} for values of $\alpha$ between 0 and 1, fixing the values of $ M = 1 $ and $ l = 0 $. 
In table~\ref{tab3} we present the comparison between the analytical and numerical results of the absorption with only quadratic GUP for values of $M$ between 0 and 1, fixing the values of $\alpha = 0.4 $ and $ l = 0 $. 
Thus, as shown in Tables~\ref{tab1},~\ref{tab2} and~\ref{tab3}, the results obtained analytically and numerically are in good agreement.
%\begin{center}
\begin{table}
\caption{ Analytical and numerical absorption results for $ \omega\rightarrow 0$ with $M=1$ and $l=0$.}
\begin{tabular}{|c|c|c|}
\hline 
$\alpha$ & Equation (\ref{abs1}) & Numerical Results \\ 
\hline 
0.0 & 16.0000 & 15.9999 \\ 
\hline 
0.1 & 15.2490 & 15.2491 \\ 
\hline 
0.3 & 14.0250 & 14.0218 \\ 
\hline 
0.5 & 13.1406 & 13.1409 \\ 
\hline 
0.7 & 12.5670 & 12.5673 \\ 
\hline 
0.9 & 12.2850 & 12.2853 \\ 
\hline 
1.0 & 12.2500 & 12.2502 \\ 
\hline
\end{tabular} 
\label{tab1}
\end{table}
%\end{center}
\begin{table}
\caption{ Analytical and numerical absorption results for $ \omega\rightarrow 0$ with $M=1$ and $l=0$.}
\begin{tabular}{|c|c|c|}
\hline 
$\alpha$ & Equation (\ref{abs2q}) & Numerical Results \\ 
\hline 
0.0 & 16.0000 & 16.0021 \\ 
\hline 
0.2 & 16.1604 & 16.1624 \\ 
\hline 
0.4 & 16.6464 & 16.6474 \\ 
\hline 
0.6 & 17.4724 & 17.4738 \\ 
\hline 
0.8 & 18.6624 & 18.6626 \\ 
\hline 
1.2 & 22.2784 & 22.2797 \\ 
\hline
\end{tabular} 
\label{tab2}
\end{table}
%\end{center}
\begin{table}
\caption{ Analytical and numerical absorption results for $ \omega\rightarrow 0$ with $\alpha=0.4$ and $l=0$.}
\begin{tabular}{|c|c|c|}
\hline 
$ M $ & Equation (\ref{abs2q}) & Numerical Results \\ 
\hline 
1.000 & 16.6464 & 16.6474 \\ 
\hline 
0.500 & 4.66560 & 4.66716 \\ 
\hline 
0.140 & 1.28013 & 1.28098 \\ 
\hline 
0.030 & 7.76551 & 7.76645 \\ 
\hline 
0.015 & 29.0880 & 29.0883 \\ 
\hline 
0.010 & 64.6416 & 64.6354 \\ 
\hline
\end{tabular}
\label{tab3} 
\end{table}
%\end{center}

\begin{figure}[htbh]
 \centering
{\includegraphics[scale=0.5]{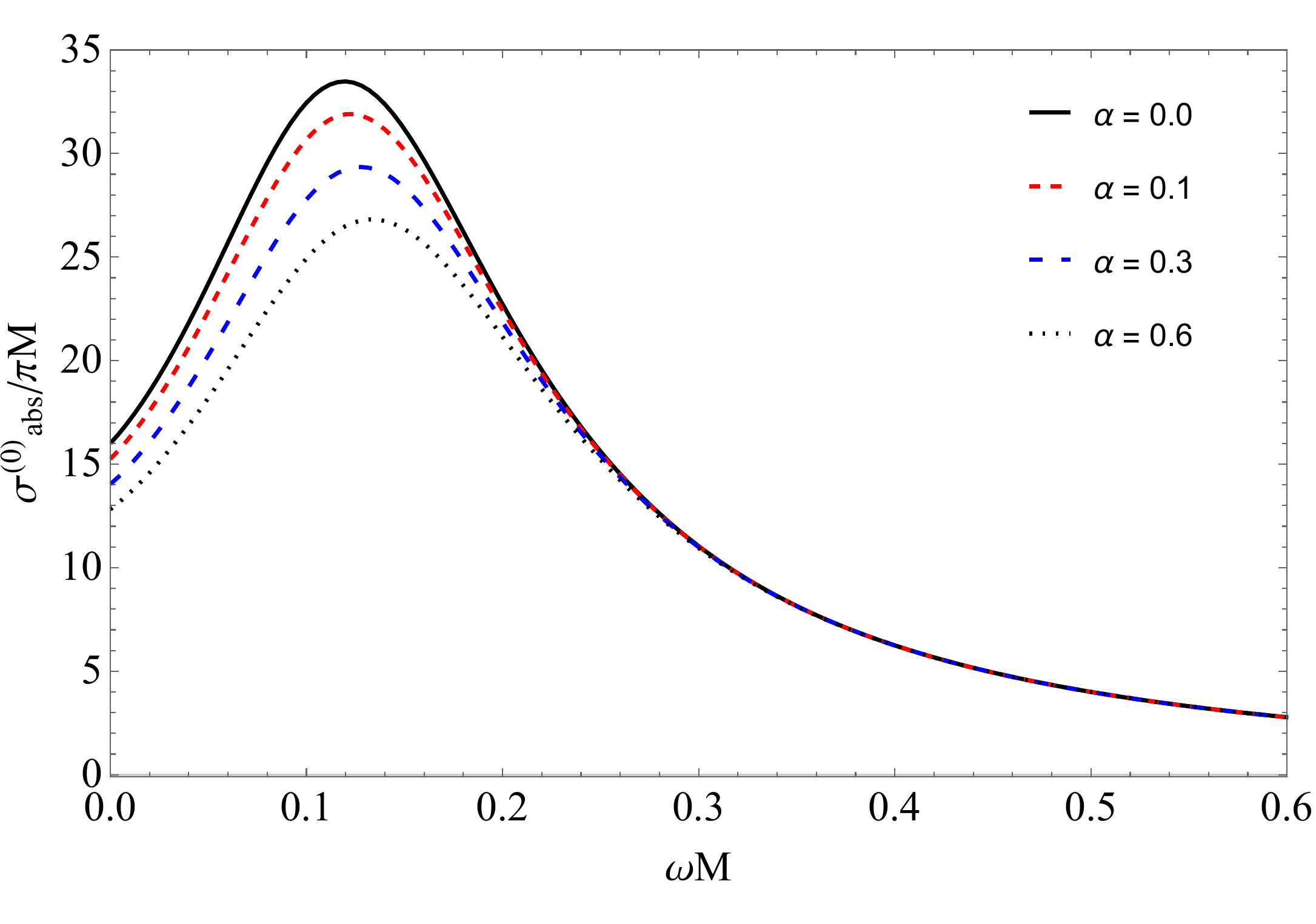}}
 \caption{Partial absorption cross section with linear and quadratic GUP for $ l=0 $, with $ M=1$  and $ \alpha=0.0, 0.1, 0.3, 0.6 $. }
\label{abslo}
\end{figure}

\begin{figure}[htbh]
 \centering
{\includegraphics[scale=0.5]{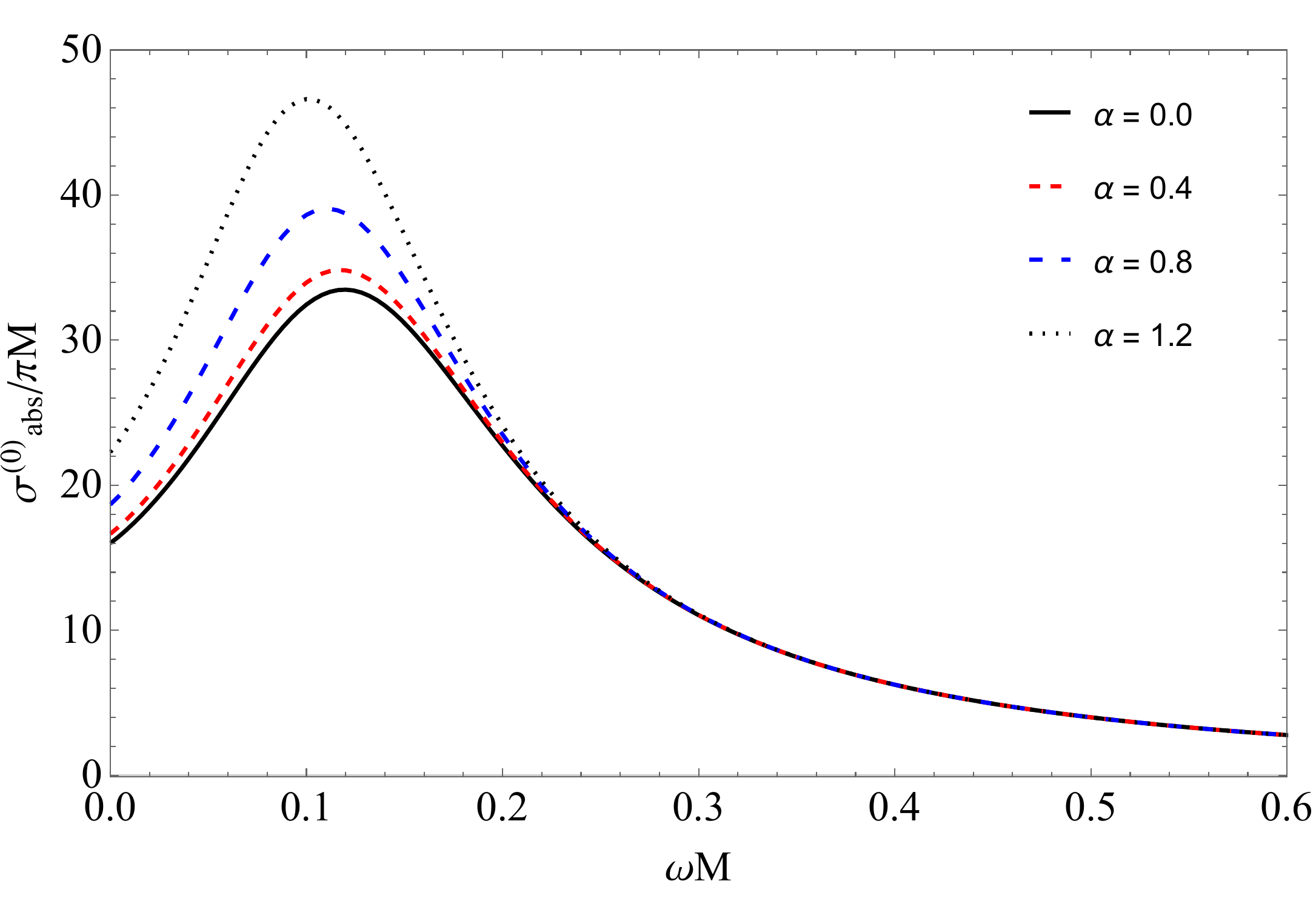}}
 \caption{Partial absorption cross section with only quadratic GUP for $ l=0 $, with $ M=1$  and $ \alpha=0.0, 0.4, 0.8, 1.2 $. }
\label{absloq}
\end{figure}

\begin{figure}[htbh]
 \centering
{\includegraphics[scale=0.5]{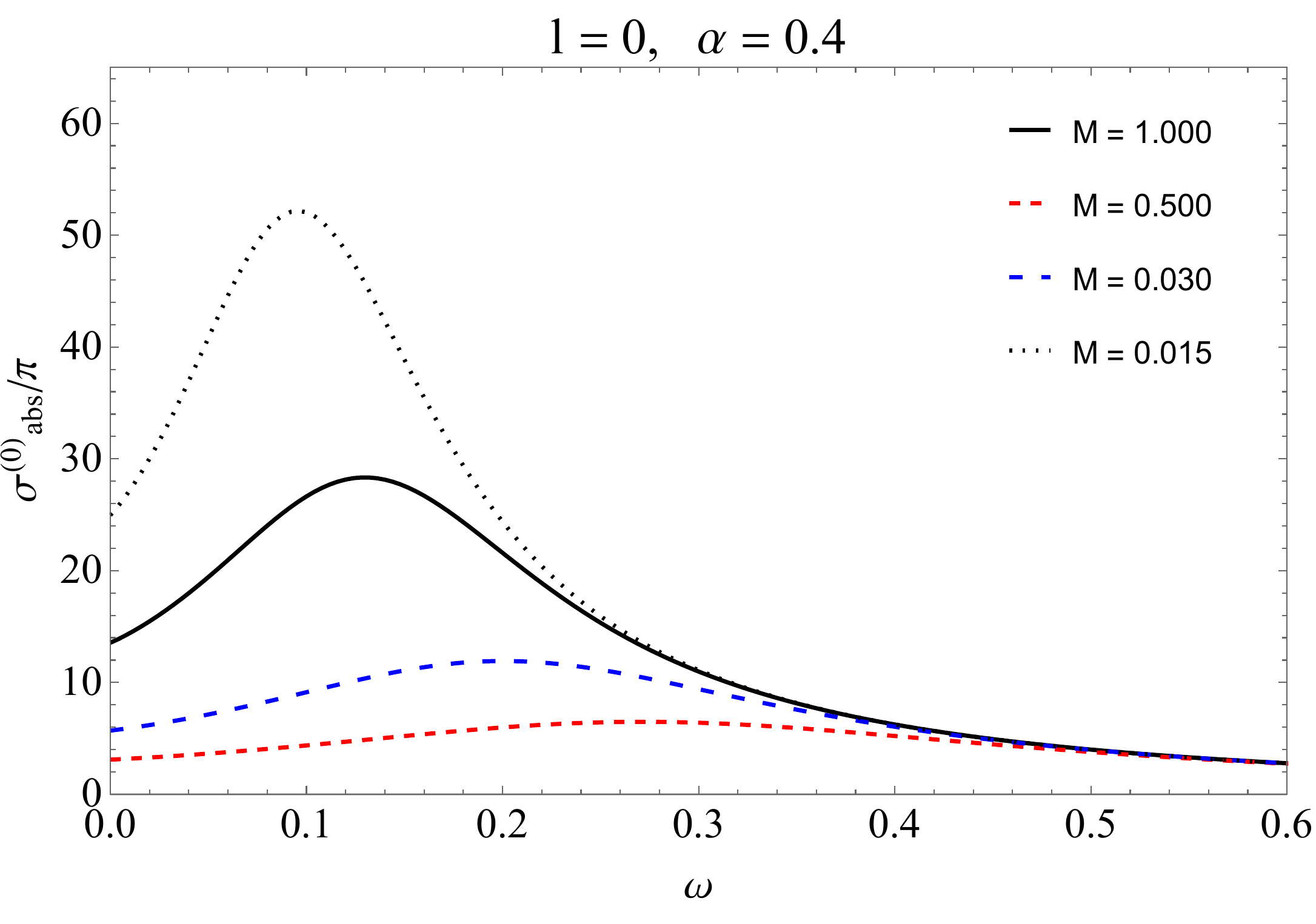}}
 \caption{Partial absorption cross section with linear and quadratic GUP for $ l=0 $, with $ \alpha=0.4$ and $ M=1.0,0.002 , 0.001$. }
\label{abslosch}
\end{figure}

\begin{figure}[htbh]
 \centering
{\includegraphics[scale=0.5]{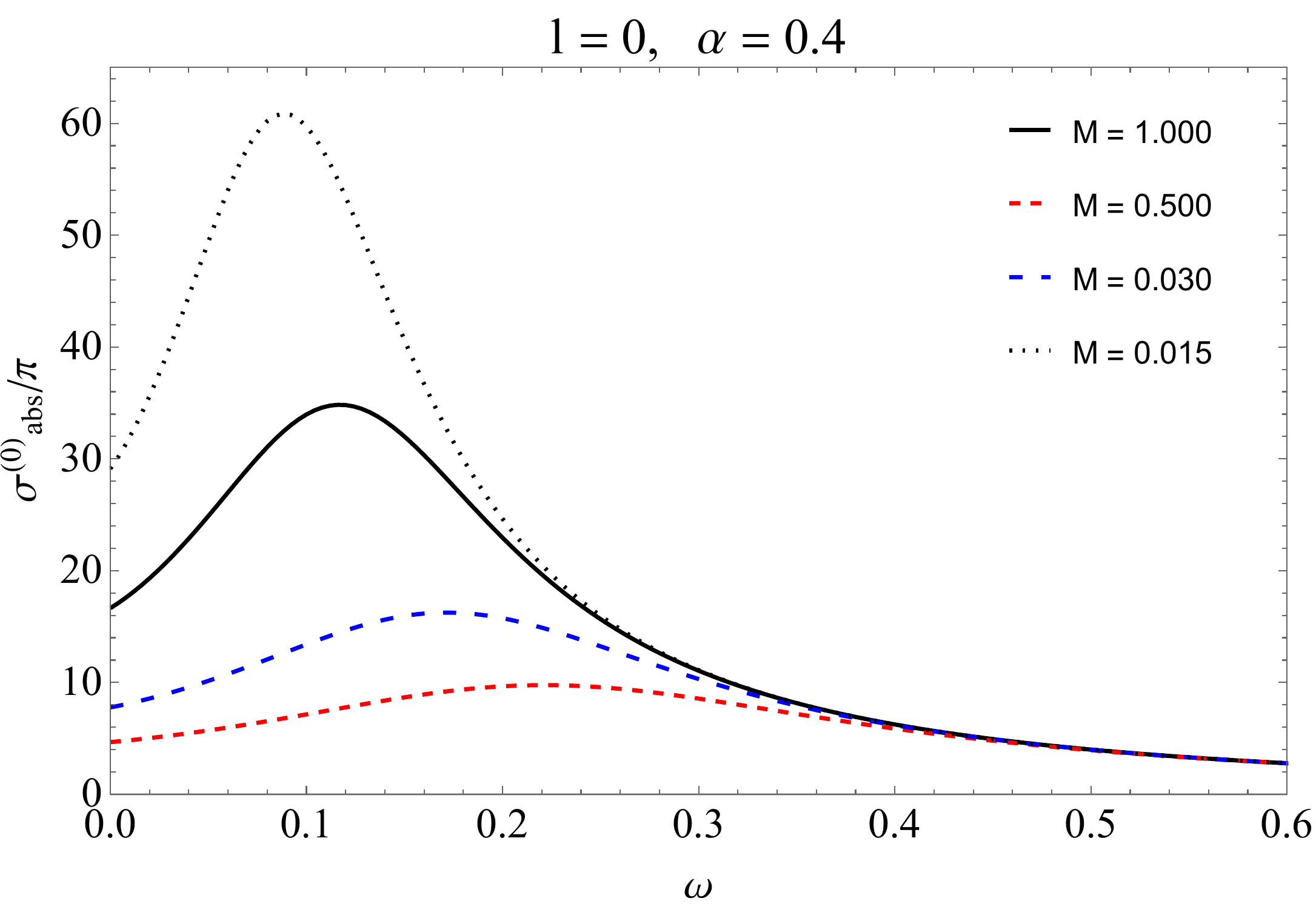}}
 \caption{Partial absorption cross section with only quadratic GUP for $ l=0 $, with $ \alpha=0.1$ and $ M=1.0,0.500, 0.030 , 0.018$. }
\label{absloschq}
\end{figure}

By considering the {\it linear and quadratic GUP}, we have plotted in Fig.~\ref{abslo} the partial absorption for mode $ l = 0 $ with $ M = 1 $,  and adopting the following values for the GUP parameter $ \alpha = 0.1, 0.3, 0.6 $.
Analyzing the curves, we find that the absorption amplitude of the Schwarzschild black hole with GUP is decreased as we vary the $ \alpha $ parameter.
In Fig.~\ref{absloq}, considering only {\it quadratic GUP}, we have plotted the partial absorption for mode $ l = 0 $ with $ M = 1 $,  and adopted the following values for the GUP parameter $ \alpha = 0.1, 0.4, 0.8, 1.2 $.
We find that the absorption amplitude is now increased when we increase the values of the parameter $\alpha$. 
By considering the {\it linear and quadratic GUP} we can see in Fig.~\ref{abslosch} that when reducing the mass value the absorption amplitude does not vanish.  
Finally by adopting only the {\it quadratic GUP}, Fig.~\ref{absloschq} shows that the absorption amplitude also does not vanish as the mass is reduced. For $ M = 0.015 $ the amplitude is greater than in the graph in Fig.~\ref{abslosch}.

\section{Conclusions}
\label{conc}
In this work we have analyzed the process of massless scalar wave scattering due to a Schwarzschild black hole with quantum corrections implemented by the GUP through the partial wave method.
We have computed the phase shift analytically at the low energy limit, and then have shown that the dominant contribution at the small angle limit of the differential scattering cross section is modified due to the GUP parameters.
We have also found that the result for the absorption cross section is given by the event horizon area of the Schwarzschild black hole with quantum corrections introduced by the GUP at the low frequency limit. 
We have also shown that, contrarily to the Schwarzschild black hole, the differential scattering/absorption cross section is nonzero at the zero mass limit.
Thus, we have found that at the limit of $ M\rightarrow 0 $ the absorption cross section presents a dominant contribution that is inversely proportional to the mass $ M $, i.e., 
$\sigma_{abs}^{\mathrm{l f}}\approx {\pi \alpha^2}/M^2$.
In addition, we have verified these results by numerically solving the radial equation for arbitrary frequencies.
It is interesting to mention that this obtained result shows similarities with the result obtained in the case of the self-dual black hole~\cite{Anacleto:2020zhp} and also with the case of the non-commutative black hole~\cite{Anacleto:2019tdj}.
For the non-commutative black hole, the absorption at the zero mass limit does not vanish and presents a result independent of the mass, that is, $\sigma_{abs}^{\mathrm{l f}}\approx {64\, \theta}$, where $\theta$ is the non-commutativity parameter.  

\acknowledgments
We would like to thank CNPq, CAPES and CNPq/PRONEX/FAPESQ-PB (Grant no. 165/2018),  for partial financial support. MAA, FAB and EP acknowledge support from CNPq (Grant nos. 306962/2018-7, 312104/2018-9, 304852/2017-1).

\end{document}